\documentclass{emulateapj}
\usepackage{lscape}

\newcommand{\lya}{Ly$\alpha$}
\newcommand{\unitcgssb}  {ergs\,s$^{-1}$\,cm$^{-2}$\,arcsec$^{-2}$}

\newcommand{\unitcgslum} {ergs\,s$^{-1}$}
\newcommand{\hh}         {$h_{70}$}
\newcommand{\hhc}        {$h_{70}^3$}
\newcommand{\hv}         {$h_{70}^{-1}$}
\newcommand{\hvc}        {$h_{70}^{-3}$}
\newcommand{\nb}         {{\sl NB}}
\newcommand{\bw}         {{\sl B}$_{\rm W}$}
\newcommand{\CIV}        {\hbox{{\rm C}\kern 0.1em{\sc iv}}}
\newcommand{\hard}       {2$-$7\,keV}

\newcommand\fion[2]{#1$\;${\scshape{#2}}}

\slugcomment{Accepted in ApJ.}

\shorttitle{Environment of Ly$\alpha$ Blobs}
\shortauthors{Yang et al.}

\begin{document}

\title{Extended L\lowercase{y$\alpha$} Nebulae at \lowercase{$z \simeq 2.3$}: \\
An Extremely Rare and Strongly Clustered Population?\,\altaffilmark{1}}

\author{Yujin Yang, Ann Zabludoff, Christy Tremonti\altaffilmark{2},
Daniel Eisenstein, Romeel Dav\'e}


\affil{Steward Observatory, University of Arizona, 933 North Cherry
Avenue, Tucson AZ 85721}
\email{yyang@as.arizona.edu}

\altaffiltext{1}{Observations reported here were obtained at the MMT
Observatory, a joint facility of the University of Arizona and the
Smithsonian Institution.}

\altaffiltext{2}{Hubble Fellow, current address: Max-Planck-Institut
f\"ur Astronomie, Heidelberg, Germany}

\begin{abstract}


To obtain an unbiased sample of bright \lya\ blobs ($L_{\rm{Ly\alpha}}$
$\gtrsim$ $10^{43\,}$\unitcgslum), we have undertaken a blind, wide-field,
narrow-band imaging survey in the NOAO Deep Wide Field Survey Bo\"otes
field with the Steward Bok-2.3m telescope.  After searching over 4.82
deg$^2$ at $z=2.3$, we discover four \lya\ blobs with $L_{\rm{Ly\alpha}}$
= 1.6 -- 5.3 $\times10^{43}$ \unitcgslum, isophotal areas of 28 --
57\,\sq\arcsec, and broad \lya\ line profiles ($\Delta v$ = 900 --
1250 km s$^{-1}$).
In contrast with the extended \lya\ halos associated with high-$z$ radio
galaxies, none of our four blobs are radio-loud.  The X-ray luminosities
and optical spectra of these blobs are diverse.  Two blobs (3 and 4) are
X-ray-detected with $L_X$(2$-$7 keV) = 2--4 $\times10^{44\,}$ \unitcgslum\
and have broad optical emission lines (\ion{C}{4}) characteristic of AGN,
implying that 50\% of our sample blobs are associated with strong AGN.
The other 50\% of blobs (1 and 2) are not X-ray or optically-detected
as AGN down to similar limits.  
The number density of the four blobs is $\sim$ 3 $\times$ 10$^{-6}$
\hhc\, Mpc$^{-3}$, comparable to that of galaxy clusters at similar
redshifts and $3\times$ lower than that found in the SSA22 proto-cluster
at $z=3.1$, even after accounting for the over-density of that region.
The two X-ray undetected blobs are separated by only 70\arcsec\ (550
kpc) and have almost identical redshifts ($\lesssim 360$ kpc along
the line-of-sight), suggesting that they are part of the same system.
Given the rarity of the blobs and our discovery of a close pair, we
speculate that blobs occupy the highest density regions and thus may be
precursors of today's rich cluster galaxies.

\end{abstract}


\section{Introduction}

The recent discovery of extended \lya\ nebulae, the so-called \lya\
blobs, may ultimately provide clues to how galaxies formed in early
Universe ($z$ $\sim$ 2 -- 5).  The most luminous \lya\ blobs have been
discovered mostly by deep narrow-band imaging surveys and have typical
sizes of $\sim$ 10\arcsec\ ($\sim$ 100 kpc) and \lya\ luminosities of
$\sim10^{44}$ \unitcgslum\, \cite[e.g.,][]{Keel99, Steidel00, Francis01,
Palunas04, Dey05, Smith&Jarvis07}.  However, what powers the blobs and
what they will evolve into remains a mystery.  To begin to answer these
questions, we need first to constrain their basic properties, including
how common they are, how they are spatially distributed, and how diverse
their possible power sources may be.

Despite intense interest, these characteristics, for even the brightest
\lya\ blobs, are poorly constrained due to the absence of an appropriate
sample.  Giant, luminous blobs \citep{Steidel00} appear to be rare
\citep{Saito06} at $z=3-5$, so an efficient, large-volume survey is
required to detect any.  Furthermore, the survey must be blind to be
representative of blob statistics and must include ancillary data (in the
X-ray, other optical bands, etc.)  to better constrain possible sources of
blob emission.  To date most \lya\ blobs have been discovered by targeting
two previously known over-dense regions \citep{Steidel00, Francis01,
Palunas04, Matsuda04}.  \citet{Matsuda04} discovered 33 additional
\lya\ blobs in the proto-cluster region at $z=3.1$ where the two giant
Steidel et al. blobs reside, and spectroscopic follow-up confirms that
the blobs are located in three filamentary structures traced by compact
\lya-emitting galaxies \citep{Matsuda05}.  \citet{Palunas04} report
that four \lya\ blobs are associated with an over-dense region of \lya\
emitters around the galaxy cluster J2142-4420 at $z =2.38$.  The first
and only study of the environment of a \lya\ blob found without any prior
knowledge of its surroundings \citep{Dey05} shows it to lie in a $3\times$
over-dense region traced by \lya-emitting galaxies \citep{Prescott08}.
Although these results suggest that bright \lya\ blobs occupy higher
density regions, we still lack an unbiased, statistical measurement of
how frequent and clustered they truly are.



The origin of the blobs' \lya\ emission is similarly murky.  Proposed
mechanisms include gravitational cooling radiation from accreting gas
\citep{Haiman00,Fardal01}, galactic superwinds driven by a starburst
\citep{Taniguchi&Shioya00}, photo-ionization by extended star
formation \citep{Matsuda07}, and a hidden AGN \citep{Haiman&Rees01}.
\citet{Smith&Jarvis07} report the discovery of a giant \lya\ blob that
appears to have cooling origin, while others suggest that blobs are
experiencing outflows \citep{Wilman05} or are static \citep{Verhamme06}.
It is also not clear what central sources the blobs harbor, e.g.,
what fraction of those central sources contain AGN?
These ambiguities arise from the optically-thick nature of the \lya\
line \cite[cf.][]{Yang06}, which clouds the interpretation of the blob
kinematics, and from the lack of multi-wavelength data, which, were it
available for a representative blob sample, could reveal the nature of
any central sources.

To obtain a unbiased sample of bright \lya\ blobs ideal for determining
their properties and, ultimately, for resolving the origin of their
extended emission, we are conducting a blind, wide-field, narrow-band
imaging survey for the brightest ($L_{\rm Ly\alpha}$ $\gtrsim$ $10^{44}$
\unitcgslum) and most extended ($A_{iso} \gtrsim 50$\,\sq\arcsec)
objects, similar to those originally found by \citet{Steidel00}.
To date, only five such blobs have been discovered among several
surveys \citep{Steidel00, Francis01, Dey05, Smith&Jarvis07}.  We select
the NOAO Deep Wide Field Survey (NDWFS) Bo\"otes and Cetus fields
\citep{Jannuzi&Dey99}, as well as the Cosmic Evolution Survey (COSMOS)
field \citep{Scoville07} as our targets, given the complementary,
multi-wavelength data available for these fields.  We choose $z=2.3$ as
our survey redshift so that the optically thin H$\alpha$ $\lambda6563$
emission line from the surrounding gas falls in a relatively clean
region of the {\sl NIR} sky spectrum.  The comparison between H$\alpha$
and \lya, as well as analysis of other lines including [\ion{O}{2}]
$\lambda3727$, [\ion{O}{3}] $\lambda5007$, and \CIV, will enable us in
future papers to study the kinematics of the surrounding gas in more
detail \citep{Dijkstra06} and to better constrain the nature of any
central sources.


In this paper, we present the initial results from our \lya\ blob
survey in the NOAO Deep Wide Field Survey (NDWFS) Bo\"otes field.
In \S\ref{sec:observation}, we describe our narrow-band imaging
survey design, observations, and selection of \lya\ blob candidates.
In \S\ref{sec:result}, we present the first results from the survey
and its spectroscopic follow-up. We also discuss the number density and
environment of the \lya\ blobs in comparison with previous studies. In
\S\ref{sec:conclusion}, we summarize the results.  Throughout this paper,
we adopt cosmological parameters: $H_0$ = 70\,\hh\,${\rm km\,s^{-1}\
Mpc^{-1}}$, $\Omega_{\rm M}=0.3$, and $\Omega_{\Lambda}=0.7$. All
magnitudes are in the AB system \citep{Oke74}.


\section {Observations}
\label{sec:observation}

\begin{figure}
\epsscale{0.9}
\epsscale{1.1}
\plotone{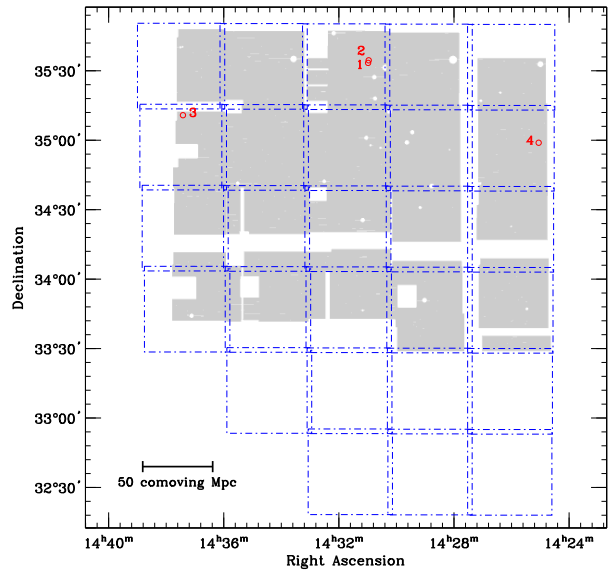}  
\caption{
Sky coverage of our narrow-band imaging survey for \lya\ blobs within the
NOAO Deep Wide Field Survey (NDWFS) Bo\"otes field.  The shaded region
represents the area covered in our Bok 2.3m + \nb403 filter survey, and
the dot-dashed lines represent the sub-fields of the NDWFS.  The gaps in
the sky coverage are due to the large gaps between the 90Prime CCDs and
a large trap in one of the CCDs.  The total sky coverage is 4.82 deg$^2$,
excluding bright stars and artifacts, making our survey one of the widest
FOV, narrow-band imaging surveys to date.  The four circles represent
the \lya\ blobs that we have discovered (see \S\ref{sec:result_multi}
for details).  None are radio-loud.  The easternmost and westernmost blobs
are detected in X-rays, while the the pair of blobs in the upper middle
region is not. 
}
\label{fig:survey_layout}
\end{figure}

\subsection{Survey Design}
\label{sec:survey_design}


Using the 90Prime one-degree field imager \citep{Williams04} mounted on
the Steward Observatory Bok 2.3m telescope on Kitt Peak, we carry out our
narrow-band imaging survey with a custom narrow-band filter (hereafter
\nb403 or \nb).  This narrow-band filter has a central wavelength
of $\lambda_c \approx 4030$\AA, designed for selecting \lya-emitting
sources at $z\approx2.3$, and a band-width of $\Delta\lambda_{\rm FWHM}
\approx 45$\AA\ that provides a line of sight depth of $\Delta z \simeq
0.037$ corresponding to 46.8 \hv\ Mpc at $z=2.3$ in the comoving frame.
The 90Prime camera consists of a mosaic of four 4k $\times$ 4k blue
sensitive CCDs with a plate scale of 0\farcs45 per pixel.  The detector
field-of-view is $\sim$ 0.8 deg$^2$ after accounting for cosmetic defects
in two of the CCDs.  This wide field-of-view is ideal for our survey as
the large area coverage compensates for the narrowness of the filter,
producing a survey comparable in volume to the largest intermediate band
filter survey \citep{Saito06}.

To date, we have finished narrow-band imaging observations with a
total sky coverage of $\sim$ 12 deg$^2$ in three fields where extensive
ancillary data sets are publicly available: the NOAO Deep Wide Field
Survey Bo\"otes and Cetus fields \cite[NDWFS;][]{Jannuzi&Dey99}
and the Cosmic Evolution Survey field \cite[COSMOS;][]{Scoville07}.
In this paper, we present the initial results from the survey within the
NDWFS Bo\"otes region centered at $\alpha$ = 14$^{\rm h}$ 32$^{\rm m}$
05\fs7, $\delta$ = +34\degr\ 16\arcmin\ 47\farcs5 (J2000).  We carried
out these observations from April to June 2007 over 16 dark nights.
Figure \ref{fig:survey_layout} shows our sky coverage overlayed with the
NDWFS field layout. After excluding regions with artifacts and bright
stars, we achieve a total sky coverage of 4.82 deg$^2$.  Although the
exposure times range from 6 to 20 hours, all the \lya\ blobs in our
final sample would have been detected even in the shallowest sub-fields
(\S\ref{sec:sample_selection}), suggesting that trading solid angle
for depth will improve constraints on the demographics of \lya\ blobs
in future surveys.  The sky coverage corresponds to a comoving survey
volume of 2.1 $\times$ $10^6$\,\hvc\,Mpc$^3$, which is $\sim$ 16$\times$
larger than the area covered by the deepest narrow-band \lya\ blob survey
to date \citep{Matsuda04}.


We reduce the data with the IRAF {\sl mscred} mosaic data reduction
package \citep{Valdes98}.  The data are corrected for cross-talk between
amplifiers, bias-subtracted, and dark-subtracted.  For flat-fielding,
we use twilight flats together with night-sky flats, which are
median-combined from object frames without the bright stars each
night. Satellite trails, CCD edges, bad pixels, and saturated pixels
are masked.  The astrometry is calibrated with the USNO-B1.0 catalog
\citep{Monet03} using the IRAF {\sl ccmap} task.  The images are scaled
using common stars in each frame and stacked to remove cosmic rays.
Finally, the images are transformed into the same world coordinate
systems as the NDWFS B$_W$ band images, which are resampled with a
coarse 90Prime pixel scale (0\farcs45 pixel$^{-1}$).  We observed 3--5
spectro-photometric standard stars in each night to derive extinction
coefficients and zero points for the \nb403 magnitudes. The major drawback
is the poor seeing, which, in the final combined images, ranges from
1\farcs5 to 2\arcsec\ with a median of 1\farcs7.  However, as we discuss
in \S\ref{sec:sample_selection}, the seeing does not impact our ability
to identify the blobs with sizes greater than 25\,\sq\arcsec.

To identify blob candidates requires that we subtract the continuum
emission within the \nb403 bandpass.  We estimate the continuum using
archival, deep, broad-band \bw\ images from the NOAO Deep Wide Field
Survey data set \cite[\bw{\sl RIJK};][]{Jannuzi&Dey99}.  The wide \bw\
filter ($\lambda_c$ $\approx$ 4135\AA\ and $\Delta \lambda$ $\approx$
1278\AA) encompasses our \nb403 filter near the central wavelength. The
5-$\sigma$ limiting magnitude of the \bw\ images is $\sim$ 26.6 mag
for a 2\arcsec\ diameter aperture.  The seeing of the NDWFS \bw\ images
ranges from 0\farcs8 to 1\farcs47 with a median of 1\farcs2.  A total
of 20 NDWFS tiles cover our narrow-band imaging fields as shown Figure
\ref{fig:survey_layout}.

%
%

\begin{figure}
\epsscale{1.0}
\epsscale{1.17}
\plotone{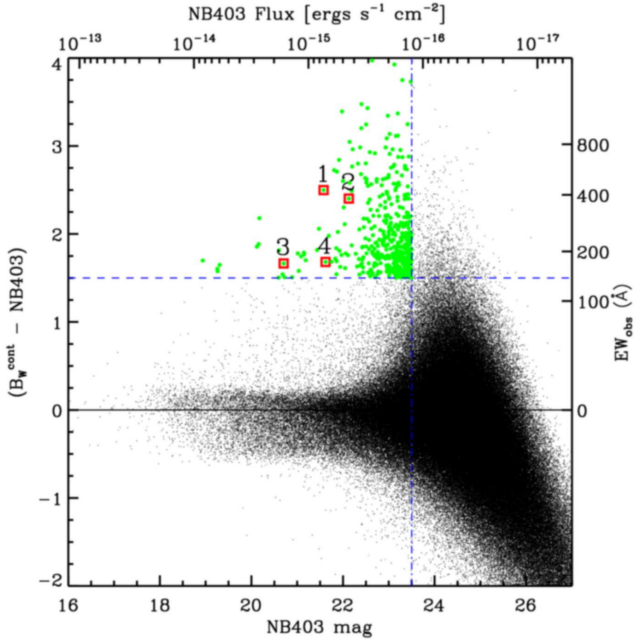}  
\caption{
The ({\sl B}$^{cont}_{\rm W}$ $-$ \nb403) color-magnitude diagram for all
sources detected in either the \bw\ and \nb\ images within the Bo\"otes
survey field. Right and top axes show the corresponding equivalent
widths in the observed frame and \nb\ fluxes, respectively.  The sharp
boundary at \nb403 $\gtrsim$ 24.0 is due to sources detected
only in the \bw\ band.  We select line-emission objects with the criteria
\nb\, $<$ 23.5 (dot-dashed line) and {\sl B}$^{cont}_{\rm W}$ $-$ \nb\
$>$ 1.5 (dashed line; $EW_{\rm obs}$ $>$ 140\AA). Squares represent the
final \lya\ blob candidates (see Fig.\ref{fig:size_luminosity}).
}

\label{fig:color_mag}
\end{figure}

\subsection{Selection of \lya\ Blob Candidates}

\label{sec:sample_selection}

To find \lya\ blob candidates, we construct photometric catalogs in
the \nb403 and \bw\ bands using SExtractor \citep{Bertin&Arnouts96}.
Because our narrow-band images are much shallower and typically have worse
seeing than the broad-band images, we make ``detection'' images (\nb+\bw)
by adding the \nb403 and \bw\ images after scaling them according to
their signal-to-noise ratios (S/N).  After identifying sources in the
``detection'' images that have least 5 pixels that are 1.5$\sigma$ above
the local sky,  we run SExtractor in double-image mode on the \nb403 and
\bw\ images with these detection images. In other words, we first find
the sources in the detection images and then obtain photometry at their
position in the \nb403\ and \bw\ images to make two separate (narrow- and
broad-band) catalogs.  We adopt Kron-like elliptical aperture magnitudes
(i.e., {\sf MAG\_AUTO} in SExtractor) to derive photometric properties.
Our use of the ``detection'' images ensures that 1) all the sources
detected in either the \nb\ or \bw\ band are included in our catalog and
2) the elliptical apertures determined by the more extended sources in
(\nb+\bw) are large enough to include all the light from both the \nb\
and \bw\ images.  This last point is critical because we do not attempt
to match the seeing between the two bands at this stage.


The selection of \lya\ blob candidates from the \nb\ and \bw\ photometry
catalogs consists of two steps: 1) selection for line (hopefully, \lya)
emitting objects with large line equivalent widths and 2) selection for
spatially extended objects with a larger angular extent in line emission
than in the broad-band.


First, we choose candidates by requiring that they are detected
above the completeness limits of the \nb403 images (\nb403  $< 23.5$
mag).\footnotemark\ We also require that candidates have observed-frame
equivalent widths larger than 140\AA\ ($EW_{\rm rest} > 42$\AA)
corresponding to ${\sl B}^{cont}_{\rm W} - \nb\ > 1.5$ (vertical and
horizontal lines in Fig. \ref{fig:color_mag}).  Candidates must be extended
(SExtractor CLASS\_STAR $>$ 0.95 in {\it both} the \nb\ and \bw\ bands).
There are a total of $\sim$ 450 objects over 4.82 deg$^2$ satisfying these
criteria.  Because most of the \lya\ blobs ($\sim$30/35) discovered by
\citet{Matsuda04} have $EW_{\rm obs}$ $\gtrsim$ 200\AA\ ($EW_{\rm rest}$
$>$ 50\AA), our selection criteria would include them if blob properties
remain the same between $z=2.3$ and $3.1$.

\footnotetext{The completeness limit varies by $\pm 0.5$ mag depending
on the sub-field due to seeing variations and different exposure
times. However, our final blob sample does not depend on the choice of
completeness limit because all the blobs are at least $\sim$ 1.5 mag brighter
than this cut (\S\ref{sec:result}).}

We estimate the continuum flux density, $f^{\lambda}_{cont}$, and
continuum-subtracted line flux, $F_{line}$, of these objects using the
following relations:
\begin{eqnarray}
f^{\lambda}_{cont} &=& \frac{F_B - F_{N\!B}}{\Delta\lambda_B - \Delta\lambda_{N\!B}} \\
\nonumber
F_{line}           &=& F_{N\!B}  - f^{\lambda}_{cont} \Delta\lambda_{N\!B}, 
\label{eq:continnum-subtraction}
\end{eqnarray}
where $F_B$ and $F_{N\!B}$ are the total flux in each filter derived
from the \bw\ and \nb403 magnitudes, respectively. $\Delta\lambda_B$
and $\Delta\lambda_{N\!B}$ represent the band-widths of the \bw\
and \nb\ filters, respectively.  Figure \ref{fig:color_mag} shows the
({\sl B}$^{cont}_{\rm W}$ $-$ \nb) color excess as a function of \nb\
magnitude for all objects detected in either the \nb\ and \bw\ bands
within our survey area.  Here, {\sl B}$^{cont}_{\rm W}$ represents the
broad-band continuum magnitude corrected for the emission line: {\sl
B}$_{\rm W}^{cont}$ = $-2.5 \log f^{\nu}_{cont} - 48.6$.


At our survey redshift, the only possible interlopers are nearby
[\ion{O}{2}] $\lambda$3727 emitters at $z\approx0.08$.  However, such
objects rarely have equivalent widths larger than 100\AA\ in the rest
frame \citep{Hogg98}.  Therefore, the contamination of our $z=2.3$
\lya\ source catalog by nearby star forming galaxies is expected to
be minimal.  Even if we reduce our selection limit to $EW_{\rm obs}$
$\gtrsim$ 65\AA, we end up with the same final \lya\ blob sample and
none of our conclusions in this paper change.


\begin{figure}
\epsscale{0.9}
\epsscale{1.1}
\vspace{-0.5cm}
\plotone{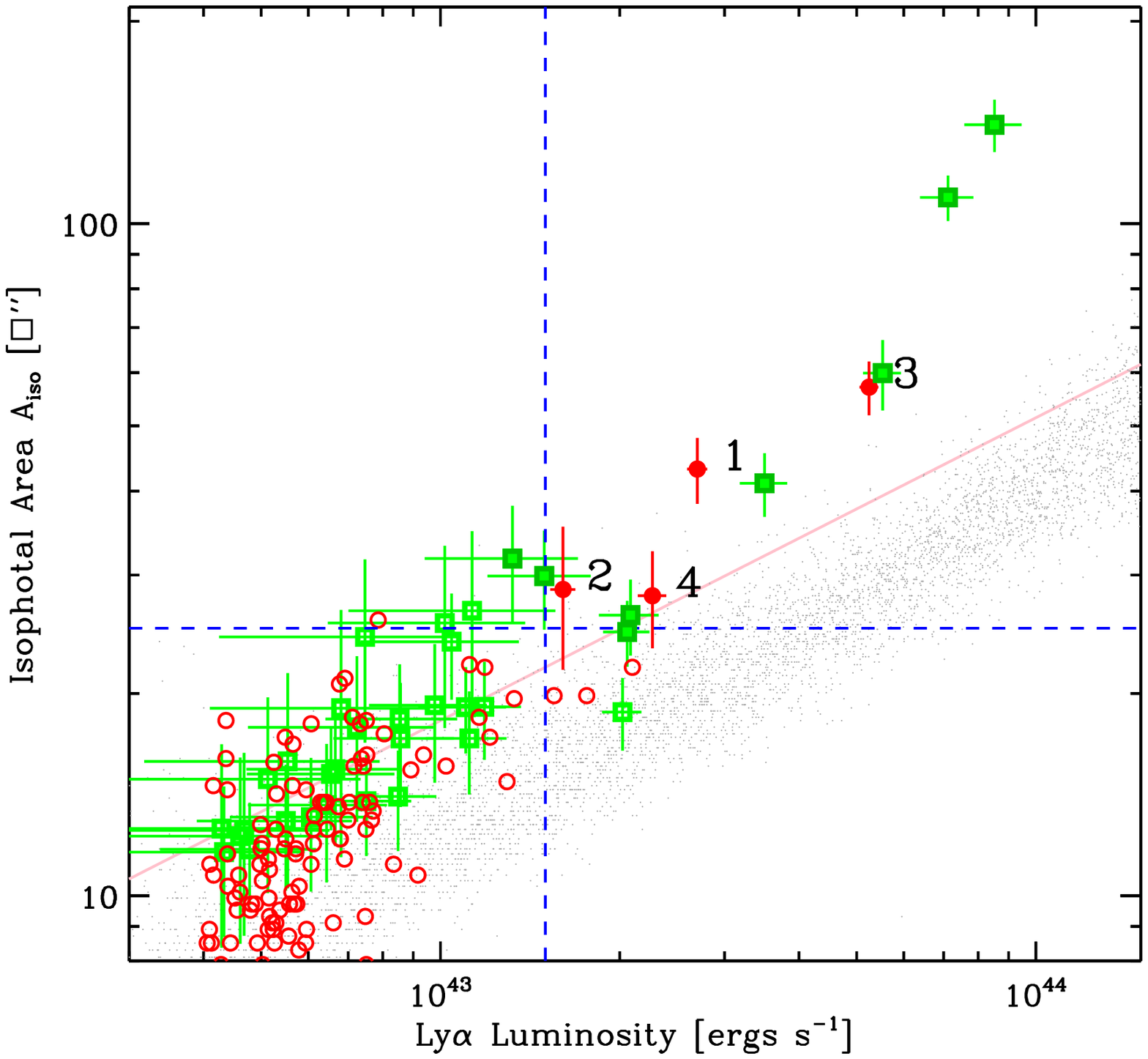}  
\caption{
Distribution of isophotal areas and \lya\ luminosities of
\lya-emitting sources (circles) selected with the first criteria
(Fig. \ref{fig:color_mag}).  The horizontal and vertical dashed
lines represent the selection criteria for the final \lya\ blob
candidate sample: $A_{iso} > 25\,\sq\arcsec$ and $L(\rm Ly\alpha) >
1.5\times10^{43\ }$\unitcgslum, respectively.  Gray dots represent the
$A_{iso}$--$L_{\rm Ly\alpha}$ relation for simulated point sources.
Note that our four final blob candidates (filled circles) are located
well above this relation (the solid line represents the upper $2\sigma$
limit), indicating that they are extended sources despite the poor seeing.
Simulated point sources mix with extended \lya-emitting sources below our
selection limits on $A_{iso}$ and $L_{\rm Ly\alpha}$, making it difficult
to distinguish blobs from point sources there.  The squares represent
the predicted sizes and \lya\ luminosities of the 35 \citet{Matsuda04}
blobs were they observed with our observational set-up at $z=2.3$.
Our final blob candidate selection criteria (dashed lines) would find
6 -- 8 (filled squares) of the brightest and largest \citet{Matsuda04}
\lya\ blobs (see \S \ref{sec:result_abundance}).
}
\label{fig:size_luminosity}
\end{figure}

Second, we identify those line-emission selected objects that are
more spatially extended in \lya\ than their continuum counterparts
(Fig.\,\ref{fig:size_luminosity}).\footnotemark\ We measure the spatial
extents of the \lya\ emission in the continuum-subtracted images.
After registering the \nb403 and \bw\ images at the sub-pixel level and
matching their seeing, we construct continuum-subtracted \nb403 images
by applying the relations in Eq.\,(\ref{eq:continnum-subtraction})
in 2--D.  We measure the isophotal area of the emission region by
running SExtractor with a threshold of $5 \times 10^{-18}$ \unitcgssb.
This measurement threshold is $\sim$ 2.3$\times$ higher than that
adopted by \citet{Matsuda04}, but because our survey redshift ($z=2.3$)
is lower than theirs ($z=3.1$), we gain a factor of $\sim$ 2.4 in surface
brightness. Although this measurement threshold is comparable to the rms
noise of the continuum-subtracted \nb403 images ($\sim$ 0.5 -- 1 $\times
10^{-17}$ \unitcgssb\ depending on the field), we are still
able to measure the sizes over this measurement limit because the bright
central parts of the objects were already detected with high S/Ns.


\footnotetext{\citet{Saito06} selected spatially extended objects by
requiring the FWHM in their intermediate-band to be larger than that in
the broad-band image, but the large seeing difference between our \nb403
images ($\sim$1\farcs7) and the NDWFS \bw\ images ($\sim$ 1\farcs2)
prevents us from adopting this approach.}

One of the potential problems in detecting a blob is the contamination of
our extended \lya\ candidate sample by point sources, especially given
the poor seeing of the narrow-band images.  To quantify this effect,
we first place artificial point-sources with a range of luminosities
($L_{\rm Ly\alpha}$ = 10$^{42}$ -- 10$^{45}$ \unitcgslum) into the sky
regions over the whole NDWFS field and measure their sizes and fluxes in
the same manner as for the extended sources.  We then determine isophotal
area and line luminosity limits above which extended and point sources
can be differentiated.

Figure \ref{fig:size_luminosity} shows the distribution of the angular
sizes and line luminosities of the 185 \lya-emitting candidates assuming
that they are all located at $z=2.3$.  The open and filled circles
represent the line-emitting objects selected using the line-emission
criteria (Fig. \ref{fig:color_mag}), and the gray dots show the relation
between the sizes and brightnesses for the artificial point sources.  As
final blob candidates, we select objects with isophotal areas larger than
25\,\sq\arcsec\ and line luminosities brighter than $1.5\times10^{43}$
\unitcgslum.  Below these limits, extended and point sources mix, and
the sizes of blobs cannot be measured reliably as explained below.


\begin{figure}[!t]
\epsscale{0.8}
\epsscale{1.15}
\plotone{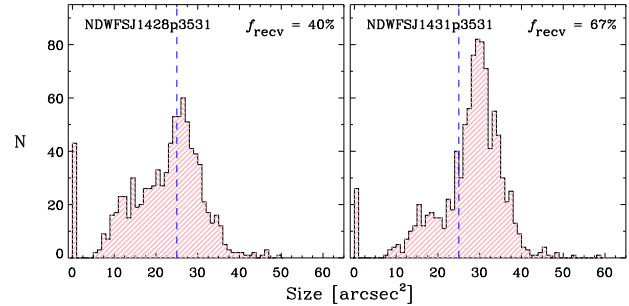}  
\caption[An example of recovery test]{
An example of recovery test for the faintest blob (blob 2) in our final
sample.  We show the distribution of the recovered isophotal area for
two NDWFS sub-fields for illustrative purposes.  This blob is actually
located in the NDWFSJ1431+3531 sub-field ({\it right}) and has the worst
recovery fraction ($f_{\rm recv}$) in the NDWFSJ1428+3531 sub-field
({\it left}). The bins with zero size indicate non-detections. The
vertical dashed line represents our selection criteria for blob size
(25\,\sq\arcsec).
}
\label{fig:recovery_test}
\end{figure}

Because the chosen isophotal threshold is comparable to the rms
sky noise, we test how reliably we can measure the spatial extent of
the blob candidates.  We cut out a small (51$\times$51 pixels) section
around each blob candidate from the (\nb\ $-$ \bw) image, filter it with
a smoothing kernel,\footnotemark\ place each postage stamp into 300 --
1000 empty sky regions in each of the 20 NDWFS sub-fields in Figure
\ref{fig:survey_layout}, and extract the sources with SExtractor in the
same way as for the real data.  Then we check how often and accurately
their sizes are recovered from these simulated images. This procedure
tests how the detectability of a given blob candidate changes across
the fields due to the seeing and exposure time variations.

\footnotetext{We use a 3$\times$3 pixel convolution mask with a FWHM =
2 pixel, which is a default kernel in SExtractor.}

Figure \ref{fig:recovery_test} shows an example of this recovery test for
the faintest blob in our final sample (blob 2, see \S\ref{sec:result}).
We show the distribution of the recovered isophotal sizes of the
artificial blob for two sub-fields. While most ($\gtrsim$95\%) of time,
the artificial blob is detected, its size is measured with a large spread
because of non-uniform sky background noise. We adopt this spread of the
recovered size distributions as the error in the blob size, $A_{\rm iso}$.

\begin{figure*}
\epsscale{0.83}
\plotone{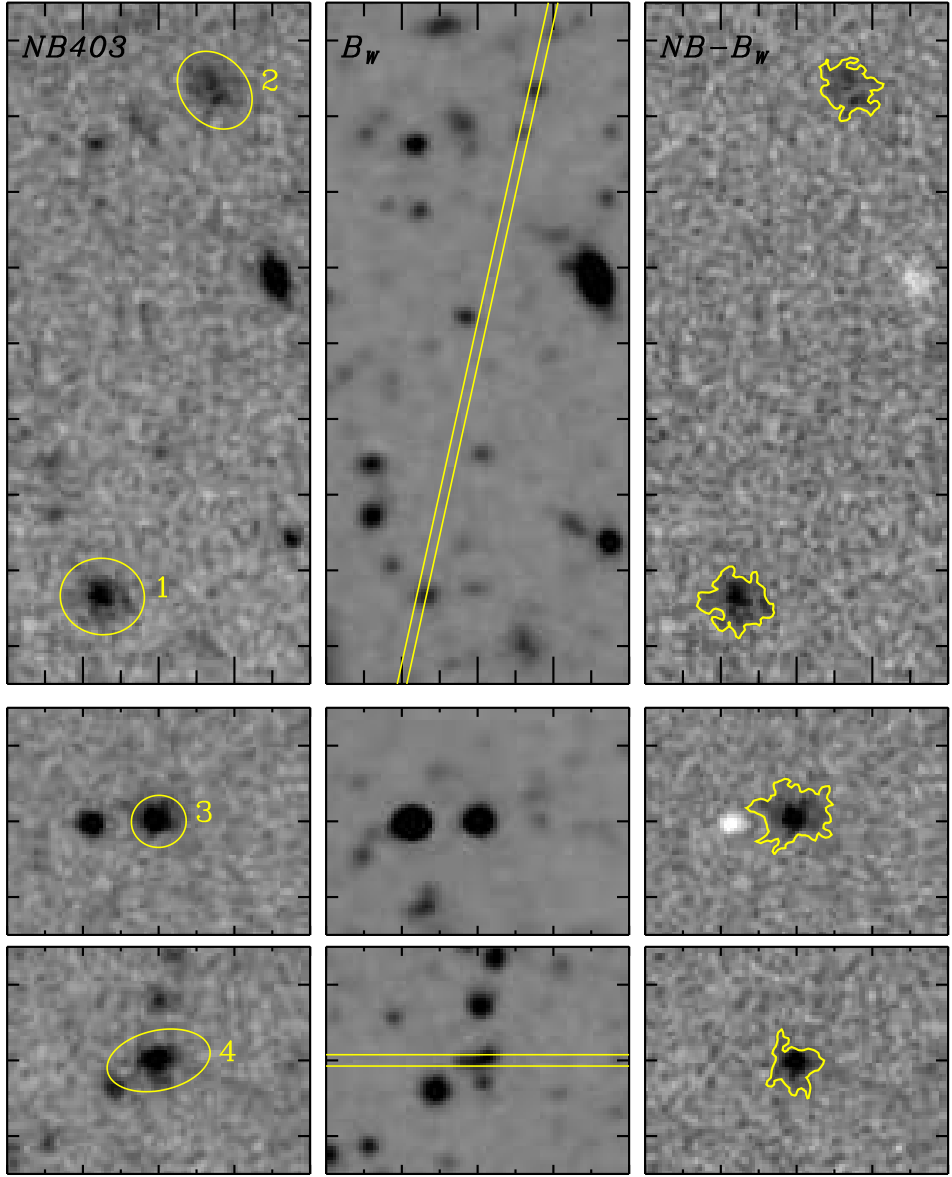}  
\caption{
Images of the four \lya\ blob candidates.  From left to right are
the \nb403, NDWFS \bw, and continuum-subtracted (\nb$-$\bw) images.
The ticks are spaced every 10\arcsec, which corresponds to $\sim$ 82
physical kpc at $z=2.3$.  The elliptical apertures used for photometry
and blob identification are shown in the left panels.  The  \bw\
images are smoothed to match the seeing of the \nb\ images.  In the
right panels, the contours represent the surface brightness limit of
$5\times10^{-18\,}$ \unitcgssb, within which we determine the blob sizes.
The location of the 1.5\arcsec-wide slit used for spectroscopic follow-up
with the 6.5m MMT is shown for blobs 1, 2, and 3 in the \bw\ image.
We spectroscopically confirm that all four blobs lie at $z=2.3$
(Fig. \ref{fig:blob_spec1d}). Blobs 3 and 4 are broad-line QSOs that
are both detected in X-rays [$L_{X}$(2--7\,keV) = 1.6 -- 4.1 $\times$
10$^{44\,}$\unitcgslum]. None of the blobs are radio-loud.
}
\label{fig:blob_image}
\end{figure*}

A recovery fraction of a blob ($f_{\rm recv}$) in a sub-field is defined
as the fraction of time that the artificial blob is recovered with a
size larger than 25\,\sq \arcsec.  These recovery fractions are averaged
over the whole field to obtain the final recovery fraction of the blob
for this survey. The size error becomes comparable to the measured size
below our sample selection criteria and the recovery fraction drops
from $\gtrsim$90\% to $\sim$50\% at the selection boundary for the
brightest blob candidates.  The recovery fractions are used to correct
the incompleteness of our survey in calculating the blob number density
in \S\ref{sec:result_abundance}.



\vspace{0.1cm}
\section{Results and Discussion}
\label{sec:result}

The selection criteria, $EW_{\rm obs}$ $>$ 140\AA, $A_{iso}$
$>$ 25\,\sq\arcsec, and $L(\rm Ly\alpha)$ $>$ 1.5$\times$ $10^{43\,}
$\unitcgslum, yield four extended \lya\ candidates at $z=2.3$ with sizes
of 28 -- 57\,\sq\arcsec, \lya\ luminosities of 1.6 -- 5.3 $\times$
10$^{43}$ \unitcgslum, and $EW_{\rm obs}$ of $\sim$ 170 -- 420\AA\
(Table \ref{tab:properties}).  Figure \ref{fig:blob_image} shows the
\nb403, \bw, and continuum-subtracted (\nb$-$\bw) images for these
candidates.  Except for Blob 4, which appears to have more than two
continuum counterparts, we are able to identify single host galaxies
in the \bw\ bands for the other three blob candidates. Although these
continuum sources are barely resolved in \bw, the line-emission regions
are extended over $\sim$ 5--10\arcsec\ in the (\nb\ $-$ \bw) images.
It is noteworthy that all four blob candidates have obvious continuum
counterparts in the deep broadband images. To date, all \lya\ blobs at
$z = 2-6$, except the blob discovered by \citet{Nilsson06}, have continuum
counterparts within the extended emission or nearby.
We also show the locations of the blob candidates on the sky in Figure
\ref{fig:survey_layout}.  Because blobs 1 and 2 are separated by only
70\arcsec\ ($\sim$ 550 physical kpc at $z=2.3$), they are shown in the
same panel in Figure \ref{fig:blob_image}.


\begin{figure*}
\epsscale{0.89}
\plotone{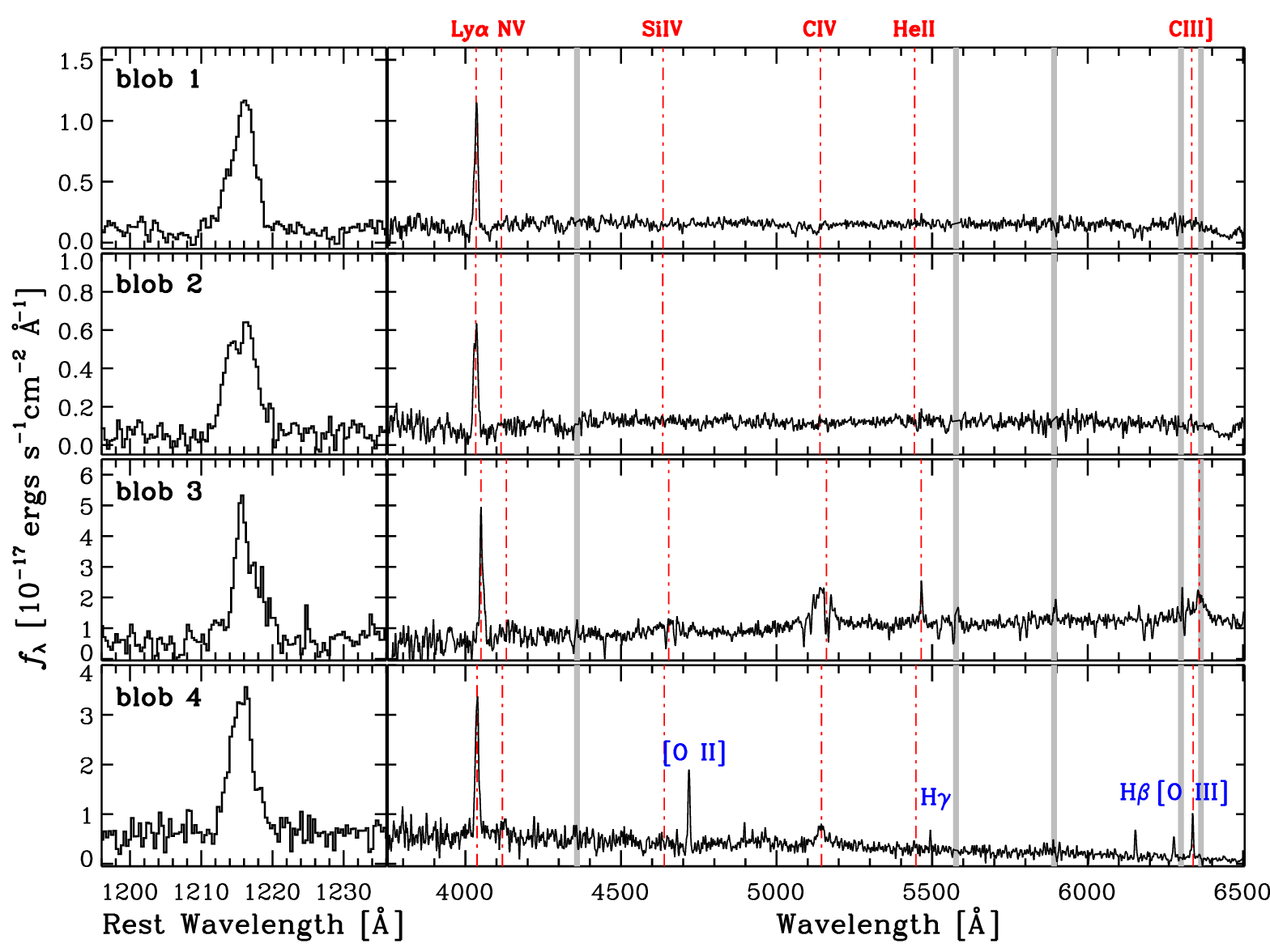}  
\caption{
Extracted 1D spectra for four blobs from 6.5m MMT long-slit spectroscopy
(blobs 1, 2, and 4) and the AGN and Galaxy Evolution Survey (blob 3;
Kochanek et al. in prep.).
For blobs 1 and 2, there is no other emission line visible up to 6500\AA.
The broad \lya\ profiles and the absence of other emission lines indicate
that blob 1 and 2 are true \lya\ blobs.  These two blobs have almost the
same redshifts ($z=2.318$) within $\delta z \simeq 0.001$.  
Their spectra do not show any AGN signatures (e.g., \CIV\,$\lambda1549$).
Neither of these blobs is X-ray-detected (\S \ref{sec:result_multi}).
Blobs 3 and 4 show other broad emission lines (e.g., \fion{C}{iv}
$\lambda1549$ and/or \fion{C}{iii}] $\lambda1909$), indicating that
these blobs host QSOs.  Unlike blobs 1 and 2, these blobs are X-ray detected
[$L_X$(\hard) = 1.6 -- 4.1 $\times$ 10$^{44}$ \unitcgslum].
Because of the blending in the \bw\ images, the blob 4 spectrum contains an
interloper, an [\fion{O}{ii}] emitter at $z\simeq0.266$ (H$\gamma$, H$\beta$,
and [\fion{O}{iii}] lines).
The shaded regions indicate where strong sky lines are present.
}
\label{fig:blob_spec1d}
\end{figure*}

\subsection{Spectroscopic Confirmation}
\label{sec:result_spectroscopy}

We spectroscopically confirm the redshifts of all four blob candidates
using spectra taken with the 6.5m ``Multiple Mirror'' Telescope (MMT) Blue
Channel Spectrograph and an existing optical spectrum from the AGN and
Galaxy Evolution Survey \cite[AGES; Kochanek et al. in prep.,][]{Cool06}.
For blobs 1, 2 and 4, we obtained low-resolution long-slit spectra on
UT 2008 February 13, April 29, and June 7.  We used the 500 gpm grating
with a 1\farcs5 $\times$ 180\arcsec\ slit, which provides a dispersion
of 1.2\AA\ per pixel and a spectral resolution of 5.4\AA\ over the
wavelength range 3400 -- 6400\AA\ (1000 -- 1900\AA\ in the rest frame).
Because blobs 1 and 2 are separated by only $\sim$ 70\arcsec, we observed
them simultaneously for a total of 6 hours with the slit position (P.A. =
$-$12.1\degr) shown in Figure \ref{fig:blob_image}.  Blob 4 is the most
ambiguous object in our sample as its continuum counterpart is not clear
due to the blending in the \bw\ image.  Therefore, we put a slit at P.A. =
90.0\degr\ to include all the continuum components seen in the \bw\
band images.  The spectra were reduced, extracted, and flux-calibrated
in the standard manner with {\tt IRAF}.


The MMT spectra confirm that blobs 1, 2 and 4 are indeed $z=2.3$
\lya\ blobs. We show the extracted spectra of these three blobs in
Figure \ref{fig:blob_spec1d}. In each case, there is an emission line
at $\simeq$\,4030\AA, which agrees well with the central wavelength of
the \nb403 filter.  For blobs 1 and 2, there are no other emission lines over
the entire wavelength range, which covers the redshifted [\ion{O}{2}]
$\lambda$3727, H$\beta$ $\lambda$4868, and [\ion{O}{3}] $\lambda$5007
lines from any $z=0.08$ star-forming galaxies.  The \lya\ lines have large
velocity widths of $\sim$ 900 and 1250 km s$^{-1}$, and observed-frame
equivalent widths of $\sim$ 150\AA\ and 190\AA, respectively.  The broad
line profile and the absence of other emission features in blobs 1 and
2 indicate that the emission line is \lya.  The stellar continuum of
blobs 1 and 2 bears a strong resemblance to the UV spectra of Lyman break
galaxies \cite[e.g.,][]{Shapley03}.  Notably, there is no evidence of
broad emission at the wavelengths of \ion{Si}{4} and \ion{C}{4},
so the presence of an unobscured AGN is ruled out with confidence.
A detailed analysis of the continuum and spectral features of these
blobs will be presented in a future paper.
For blob 4, we identify two continuum sources, an [\ion{O}{2}] emitter
at $z\simeq0.266$ and a QSO that produces broad \lya\ and \ion{C}{4}
lines at $z\simeq2.32$. The \lya\ line has a velocity width of $\sim$
950 km s$^{-1}$ and an observed-frame equivalent width of $\sim$ 85\AA.
Note that the EWs measured from the long-slit spectra are lower than
those estimated from the narrow-band imaging due to slit-loss.

For blob 3, we obtain a spectrum from AGES, which has extensive optical
spectroscopy for all the X-ray detected galaxies with {\sl I}$_{\rm
AB}$ $\lesssim$ 22.0 mag in the Bo\"otes field.  This spectrum
shows that blob 3 is at $z=2.32$ and contains a QSO with strong,
broad \lya\ emission (Figure \ref{fig:blob_spec1d}).  Blob 3 is
perhaps similar to the diffuse \lya\ halos associated with some QSOs
\cite[e.g.,][]{Bunker03,Weidinger05}.

\subsection{Multi-Wavelength Properties of Confirmed \lya\ Blobs}
\label{sec:result_multi}

Are any of the four blobs associated with strong radio or X-ray sources?
None is detected in the VLA FIRST survey \citep{White97} at 1.4 GHz
at the detection limit of 1 mJy.  Assuming a power-law spectral energy
distribution of $S(\nu)$ $\propto$ $\nu^{-0.8}$, this limit corresponds
to a rest-frame 1.4 GHz luminosity density of 3.2$\times$$10^{32}$
\unitcgslum Hz$^{-1}$, which is $\sim$ two orders of magnitude
fainter than the powerful radio galaxies associated with \lya\ halos
\cite[e.g.,][]{Reuland03}.  However, in X-rays, the properties of
these blobs vary wildly.  Blobs 3 and 4, whose optical spectra have
broad QSO lines, are detected in the {\sl Chandra} XBo\"otes survey
\citep{Kenter05,Brand06} with $L$(2--7\,keV) = 1.6 -- 4.1 $\times$
10$^{44\,}$\unitcgslum.  Blobs 1 and 2 are not detected at a similar
depth and have upper limits of $L$(2--7\,keV) $<$ 0.63 and 0.18 $\times$
10$^{44\,}$\unitcgslum, respectively.  Therefore, at least two of the
blobs (50\%) are associated with AGN.  Table \ref{tab:properties} lists
the optical and X-ray properties of the four blobs.


It is unknown what powers the copious \lya\ emission of blobs. Previous
observational studies suggest diverse mechanisms, e.g., a central
AGN \citep{Dey05}, galactic superwind \cite[e.g.,][]{Wilman05},
extended star formation \citep{Matsuda07}, and cooling radiation
\citep{Nilsson06,Smith&Jarvis07,Smith08}, yet there is no smoking gun.
Due to its blind strategy, large volume, follow-up optical spectroscopy,
and overlap with wide {\sl Chandra} X-ray imaging coverage, our work
here is the first to place unbiased limits on the fraction of blobs with
luminous AGN (50\%).  The {\it diversity} of the blob optical and X-ray
properties is also interesting, suggesting that AGN like those detected
in half the blobs are not necessary to power the others.


\subsection{Rarity of \lya\ Blobs}
\label{sec:result_abundance}

Our survey indicates that extended \lya\ nebulae are extremely rare
at a redshift of $2.3$.  Based on the completeness (recovery) test in
\S\ref{sec:sample_selection}, we estimate the number of \lya\ blobs
within the survey volume as $N = \sum_{i} 1/f^{i}_{\rm recv}$, where
$f^{i}_{\rm recv}$ is the recovery fraction (Table \ref{tab:properties}).
The discovery of only four \lya\ blobs over 4.82 deg$^2$ yields a number
density of $2.5 \pm 1.1$ $\times$ 10$^{-6}$ \hhc\,Mpc$^{-3}$ for blobs
with $A_{iso} > 25\,\sq\arcsec$ and $L(\rm Ly\alpha) > 1.5\times10^{43\
}$\unitcgslum.  Note that this number density is comparable to those of
galaxy clusters in the nearby and high-$z$ Universe, $n \sim 10^{-5} -
10^{-6}$ \hhc\,Mpc$^{-3} $\cite[e.g.,][]{Bahcall03,Papovich08}. Because
our survey fails to find blobs as bright ($L_{\rm Ly\alpha}$ $\gtrsim$
$10^{44}$ \unitcgslum) and large ($A_{iso} \gtrsim 150$ \sq\arcsec) as
the brightest and largest in the \citet{Steidel00} sample (their Blobs
1 and 2), we conclude that the \citet{Steidel00} blobs are even rarer
objects with a number density of $n$ $\lesssim$ 0.5 $\times$ 10$^{-6}$
\hhc\,Mpc$^{-3}$.

To compare our blob number density with those from previous surveys
\citep{Matsuda04,Saito06}, we determine how many of the \citet{Matsuda04}
blobs would be detected in our survey if they were located at $z=2.3$.
Using the continuum subtracted \nb$_{\rm corr}$ images of the 35 blobs
from \citet{Matsuda04}, we scale the blob surface brightnesses and sizes
to $z=2.3$ according to the adopted cosmology and assuming that their
physical sizes and \lya\ luminosities do not change from $z=3.1$ to $2.3$.
These images are convolved with Gaussian kernels to match our poor seeing,
rebinned to the 90Prime pixel scale, and given Poisson noise.  We do
not account for the difference between the filter (\nb403 vs. \nb497)
band-widths, because the resolving power of two filters is similar.
We place the simulated images into our continuum-subtracted \nb403 images
and measure their sizes and luminosities in the same way as described
in \S \ref{sec:sample_selection}. We repeat this process 500--1000 times
to derive the range of recovered luminosities and sizes.

We show the distribution of sizes and \lya\ luminosities ($A_{iso}$
-- $L_{\rm Ly\alpha}$) in Figure \ref{fig:size_luminosity}.
Based on this test, we expect that 6 -- 8 of the 35 \citet{Matsuda04}
blobs\footnotemark\ are detectable given our luminosity-size criteria,
yielding an effective blob number density of 4.6--6.2 $\times$10$^{-5}$
\hhc\,Mpc$^{-3}$.  Considering the recovery rate of each Matsuda blob,
which is $\sim$100\% for $L$(\lya) $>$ 5$\times$ 10$^{43}$ and drops to
$\sim$ 50\% at our selection boundary, we expect to detect on average
$\sim$ 6.0 of their blobs with our survey criteria.  Once we factor in
the difference between our survey volume and Matsuda et al.'s, we should
have detected $\sim$98 blobs --- not four --- provided that the blob
number density at $z=3.1$ and $2.3$, and between the two survey fields,
is the same.

\footnotetext{In the detectability simulation, the average values of
extracted $A_{iso}$ and $L_{\rm Ly\alpha}$ for six \citet{Matsuda04}
blobs (their LAB 1, 2, 3, 4, 8, 10) satisfy our selection criteria.
Two other blobs (their LAB 6 and 15) are located on our selection
boundary, but satisfy the selection criteria more than 50\% of the time
in the simulation.}

The comparison of our results with those of \citet{Saito06} is more
difficult because we do not have high S/N measurements of their blob
surface brightness profiles.  Therefore, we estimate the blob number
density for $L$(\lya) $\gtrsim$ 1.5 $\times$ $10^{43\,}$\unitcgslum\
from the brightest bin in their Figure 13.  We consider the resulting
number density of $\sim$ 6.7 $\times$ 10$^{-6}$ \hhc\,Mpc$^{-3}$ an upper
limit at $z = 3$--5, given that some fraction of their blobs might not
have been detected with our survey criteria.


\begin{figure}
\epsscale{1.00}
\epsscale{1.15}
\plotone{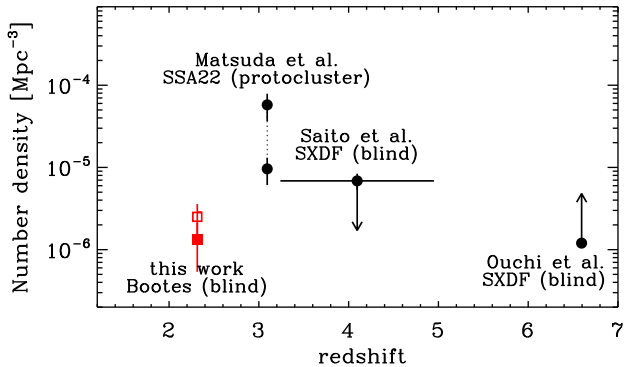}  
\caption{
Number density of \lya\ blobs at different redshifts. Open and filled
squares represent the number density from our narrow-band imaging
survey at $z=2.3$ for the blobs with and without X-ray detections,
respectively.  We also show two number density estimates for the blobs
in \citet{Matsuda04}, with or without correction for the over-density of
the SSA22 proto-cluster region. We incorporate the same luminosity-size
limits on Matsuda et al.'s result as obtained in this paper for a
proper comparison.  For blobs at higher redshifts, we show only upper
\cite[$3<z<5$;][]{Saito06} and lower limits \cite[$z=6.595$;][]{Ouchi08}
from the \lya\ blob searches in the Subaru/XMM-Newton Deep Survey field
(SXDF).
}
\label{fig:number_density}
\end{figure}

We show the number density of \lya\ blobs from the different samples and
at different redshifts in Figure \ref{fig:number_density} after applying
the same luminosity--size limits as our survey.  Open and filled squares
represent the number density estimates from our survey with and without
the two X-ray detected blobs. For the Matsuda et al. sample, we show
both the measured number density and that scaled down to compensate
for the known over-density of the SSA22 field \cite[$\delta$ $\sim$
6;][]{Steidel00}.  For Saito et al.'s survey, we show the upper-limit
discussed above and the redshift range (horizontal error bar).  We also
show a lower limit on the number density at $z=6.595$ derived from the
recent discovery of a \lya\ blob in the Subaru/XMM-Newton Deep Survey
field \citep{Ouchi08}.


At face value, \lya\ blobs within the $z=3.1$ proto-cluster from the
\citet{Matsuda04} survey are $\sim$ 20 -- 30 times more numerous than
blobs in our blind field survey.  Even if we account for the galaxy
over-density in the proto-cluster, the blob number density in our survey
is still by a factor of 3 lower, suggesting that extended \lya\ emission
is closely related to (or enhanced by) a clustered environment. However,
we cannot rule out that this discrepancy might reflect real evolution
between $z=2.3$ and $3.1$.

\subsection{A Close Pair of \lya\ Blobs}

\label{sec:clustering}

One of our most interesting results is the discovery of the pair of blobs
(1 and 2) separated by only 70\arcsec, which corresponds to $\sim$550
physical kpc at $z=2.3$.
To estimate how unlikely it is to find two blobs within 70\arcsec\ of
each other, we calculate the expected number of galaxy pairs assuming
that the spatial distribution of \lya\ blobs can be represented by a
two-point correlation function, $\xi (r,z) = [r/r_0(z)]^{-\gamma}$, where
$r_0(z)$ is the scale length of galaxy clustering at the redshift $z$.
By projecting this correlation function onto the sky \citep{Limber53},
we derive the angular correlation function $\omega(\theta)$, which
measures the excess probability above random of finding a galaxy at an
angle $\theta$ from another galaxy.
For a given power-law correlation function, one can derive $\omega(\theta)
= A_{\omega} \theta^{1-\gamma}$ \citep{Peebles80,Efstathiou91} such that
\begin{equation}
A_\omega = H_\gamma \int r^\gamma_0(z)\, d^{1-\gamma}_C(z) \left[\frac{dN}{dz}\right]^2 \frac{H(z)}{c} dz
                                                           \left[\int\frac{dN}{dz} dz\right]^{-2},
\end{equation}
where $d_C(z)$ is comoving distance at $z$, $dN/dz$ is the number of
galaxies per unit redshift interval, $H(z)$ is the Hubble parameter at
$z$, and $H_\gamma = \sqrt{\pi}\Gamma[(\gamma-1)/2]/\Gamma(\gamma/2)$.
We adopt $dN/dz$ as a top-hat function for our survey redshift interval
and assume a fiducial value of $\gamma = 1.8$.

From this angular correlation function, we estimate the number of galaxy
pairs within $\theta_p$ using:
\begin{equation}
N_p (<\theta_p) = \int_0^{\theta_p} \frac{1}{2}N(N-1)[1+\omega(\theta)]
                  \frac{2\pi\theta d\theta}{\Omega},
\end{equation}
where $\Omega$ is the survey area, and $N$ is the number of blobs found
in our survey.
For the typical galaxy correlation length, $r_0$ $\sim$ 7 Mpc in the
comoving frame \cite[e.g.,][]{Maddox90}, we predict a negligible pair
count, $N_p$ $\simeq$ 0.005.  The expected number of close pairs does not
increase dramatically as the correlation length increases. Even for the
correlation length of the richest galaxy clusters, $r_0$ $\sim$ 30 Mpc
\cite[e.g.,][]{Bahcall03,Papovich08} for redshifts out to $z \sim 1.5$,
only $N_p$ $\simeq$ 0.05 pairs are expected. Therefore, we conclude
that the observed close pair of \lya\ blobs is unlikely to occur for a
reasonable range of clustering strengths if we assume that the two \lya\
blobs belong to individual dark matter halos.




Increasing the size of our blob sample (Yang et al. 2008, in prep.) will
better constrain the clustering of \lya\ blobs. For the time being, it is
intriguing that we do discover a close pair.  Our MMT
spectra confirm that the two blobs have almost the same redshifts within
$\delta z$ $\simeq$ 0.001 and a corresponding line-of-sight separation
of $\delta r$ $\simeq$ 360 physical kpc assuming no peculiar velocity.
The similarity of their redshifts, as well as their small separation on
the sky (corresponding to 550 physical kpc), leads us to speculate that
these two blobs might lie within a single dark matter halo, i.e., within
a massive proto-group or cluster of galaxies.  
Note that the analytic $\Lambda$CDM model predicts a virial radius
($r_{200}$) of 430 kpc at $z=2.3$ for a dark matter halo with a mass
of $10^{14}$\,$M_{\odot}$ \citep{Mo&White02}.
Previous studies also suggest that \lya\ blobs are phenomena that favor
dense environments \citep{Matsuda05,Prescott08}. This hypothesis could be
tested by using a deeper narrow-band imaging survey to characterize the
environment of this blob pair with the spatial distribution of fainter,
smaller blobs and/or faint, compact \lya\ emitters.



\section{Conclusion}
\label{sec:conclusion}

In this paper, we present initial results from our blind, wide-field,
narrow-band imaging survey in the NOAO Deep Wide Field Survey Bo\"otes
field to constrain the number density, environment, and multi-wavelength
properties of extended \lya\ nebulae (``\lya\ blobs") at $z=2.3$.
After searching over 4.82 deg$^2$, we discover four \lya\ blobs with
$L_{\rm{Ly\alpha}}$ = 1.6 -- 5.3 $\times10^{43}$ \unitcgslum, isophotal
areas of 28 -- 57\,\sq\arcsec, and broad \lya\ line profiles ($\Delta v$
= 900 -- 1250 km s$^{-1}$).  We confirm the redshifts of all four blobs
spectroscopically.
In contrast with the extended \lya\ halos associated with high-$z$ radio
galaxies, none of our four blobs are radio-loud.  The X-ray luminosities
and optical spectra of these blobs are diverse.  Two blobs (3 and 4)
are X-ray-detected with $L_X$(2$-$7 keV) = 2--4 $\times10^{44\,}$
\unitcgslum\ and have broad optical emission lines (\ion{C}{4} and
\ion{C}{3}]) characteristic of AGN, implying that 50\% of our sample
blobs are associated with strong AGN.
The other 50\% of blobs (1 and 2) are not X-ray or optically-detected
as AGN down to similar limits, suggesting that AGN like those in blobs
3 and 4 are not necessary to power them.  The number density of the four
blobs is $\sim$ 3 $\times$ 10$^{-6}$ \hhc\,Mpc$^{-3}$, comparable to that
of galaxy clusters at similar redshifts and $3\times$ lower than that
found in the SSA22 proto-cluster at $z=3.1$, even after accounting for
the over-density of that region.  The two X-ray undetected blobs (1 and
2) are separated by only 70\arcsec\ (550 physical kpc) and have almost
identical redshifts (corresponding to $\lesssim$ 360 physical kpc along
the line-of-sight), suggesting that they are part of the same system.
Given the rarity of the blobs and our discovery of a close pair, we
speculate that blobs occupy the highest density regions and thus may be
precursors of today's rich cluster galaxies.

\acknowledgments

We thank the referee, William Keel, for the thorough reading of the
manuscript and helpful comments.
We thank Ed Olszewski and mountain staffs in Steward Bok 2.3m telescope
for their helps with 90Prime observing runs.
We thank Yuichi Matsuda for providing us the narrow-band images of their
\lya\ blobs.
We also thank Masami Ouchi for the helpful discussions and allowing us
to use their blob number density before the publication.
Y.\,Y. thanks Hee-Jong Seo, Suresh Sivanandam, Richard Cool, and Wiphu
Rujopakarn for their helps in clustering analysis, Chandra data, and
obtaining AGES data, respectively.
Y.\,Y. thanks Toshihiko Kimura in Asahi Spectra for his thorough work in
manufacturing NB403 filter.
%
%
Support for C. A. T. was provided by NASA through Hubble Fellowship grants
HST-HF-01192.01 awarded by the Space Telescope Science Institute, which
is operated by the Association of Universities for Research in Astronomy,
Inc., for NASA, under contract NAS5-26555.
This work made use of images and/or data products provided by the NOAO
Deep Wide-Field Survey, which is supported by the National Optical
Astronomy Observatory (NOAO). NOAO is operated by AURA, Inc., under a
cooperative agreement with the National Science Foundation.

Facilities: \facility{MMT (Blue Channel), Steward Bok2.3m (90Prime)}


\clearpage

\begin{landscape}



\newcommand\ff[1]{\tablenotemark{#1}}
\begin{deluxetable}{cccccccc cccccc}
\tablewidth{0pt}
\tabletypesize{\small}
\tabletypesize{\scriptsize}
\tablecaption{Properties of \lya\ Blobs \label{tab:blob}}
\tablehead{
\colhead{ID}&
\colhead{R.A.}&
\colhead{Dec.}&
\colhead{$z$}&
\colhead{$L$(\lya)}&
\colhead{Size}&
\colhead{$f_{\rm recv}$}&
\colhead{$EW_{\rm obs}$}&
\colhead{$L$(2-7keV)\tablenotemark{a}}&
\colhead{NDWFS Name}&
\multicolumn{4}{c}{NDWFS magnitudes\tablenotemark{b}}\\
\colhead{}&
\colhead{(J2000)}&
\colhead{(J2000)}&
\colhead{}&
\colhead{($10^{43}$\unitcgslum)}&
\colhead{(\sq\arcsec)}&
\colhead{}&
\colhead{(\AA)}&
\colhead{($10^{44}$\unitcgslum)}&
\colhead{}&
\colhead{\sl Bw}&
\colhead{\sl R}&
\colhead{\sl I}&
\colhead{\sl K}
}
\startdata
Blob 1 &   14 30 59.0 &   +35 33 24.7 &  2.3186   &       2.70 $\pm$ 0.10  &      43 $\pm$  4.8  &   0.93   &   422 $\pm$    20.8 & $\lesssim$  0.63\ff{c}        &  NDWFS J143059.0+353324      &    23.63  &   22.91  &  22.69  & \nodata \\
Blob 2 &   14 30 57.8 &   +35 34 31.6 &  2.3178   &       1.61 $\pm$ 0.08  &      29 $\pm$  6.9  &   0.57   &   383 $\pm$    22.2 & $\lesssim$  0.18\ff{c}        &  NDWFS J143057.8+353431      &    24.05  &   22.92  &  22.53  & \nodata \\
Blob 3 &   14 37 25.1 &   +35 10 48.8 &  2.3321   &       5.25 $\pm$ 0.19  &      57 $\pm$  5.2  &   0.97   &   171 $\pm$ \phn7.3 &             4.11  $\pm$  1.86 &  NDWFS J143725.0+351048      &    22.11  &   20.67  &  19.93  &   16.74 \\
Blob 4 &   14 25 03.5 &   +34 58 55.1 &  2.3211   &       2.27 $\pm$ 0.12  &      28 $\pm$  4.6  &   0.67   &   174 $\pm$    10.4 &             1.63  $\pm$  1.40 &  NDWFS J142503.4+345854\ff{d}&    23.06  &   23.70  &  22.13  & \nodata \\
       &              &               &           &                        &                     &          &                     &                               &  NDWFS J142503.3+345855\ff{d}&  \nodata  &   23.66  &  23.24  & \nodata
\enddata
\tablenotetext{a}{X-ray luminosities are derived from the observed flux
in the 0.5--2 keV band assuming a power-law spectrum with $\Gamma = 1.7$
and a Galactic absorption $N_{\rm H} = 1\times10^{20} {\rm cm}^{-1}$. \\
} 

\tablenotetext{b}{Vega magnitudes from the NDWFS DR3 catalog. \\} 

\tablenotetext{c}{The upper limits are derived from the combined Chandra
ACIS-I images (Obs-ID 3608,6993).\\}

\tablenotetext{d}{It is not clear which of two galaxies is associated
with the blob 4 and the X-ray source.\\}

\label{tab:properties}
\end{deluxetable}


\clearpage
\end{landscape}

\clearpage



\end{document}